\documentclass[%
 reprint,
superscriptaddress,
 amsmath,amssymb,
 aps, physrev, prl, longbibliography
]{revtex4-2}

\usepackage{graphicx}% Include figure files
\usepackage{dcolumn}% Align table columns on decimal point
\usepackage{bm}% bold math
\usepackage{natbib}
\usepackage{bm,bbm}% bold math
\usepackage{upgreek}
\begin{document}

\nocite{apsrev42Control}

\preprint{}

\title{\textbf{Unveiling Structured Optical Coherence in Nonlinear Optics} 
}% 

\author{Zihao Pang}
 \email{Contact author: its.zihaopang@gmail.com}

\affiliation{%
School of Electrical and Computer Engineering, Iby and Aladar Fleischman Faculty of Engineering, Tel Aviv University, Tel Aviv 6997801, Israel.
}%
\author{Ady Arie}%
 \email{Contact author: ady@tauex.tau.ac.il}

\affiliation{%
School of Electrical and Computer Engineering, Iby and Aladar Fleischman Faculty of Engineering, Tel Aviv University, Tel Aviv 6997801, Israel.
}%

\affiliation{%
Jan Koum Center for Nanoscience and Nanotechnology, Tel Aviv University, Tel Aviv 6997801, Israel.
}%

%\affiliation{
% First affiliation for this author
%}%
%\affiliation{
% second institution for this author
%\affiliation{%

\date{\today}% It is always \today, today,

\begin{abstract}
Nonlinear optical interactions are typically driven by coherent sources. Recently, however, incoherent sources have also been shown to offer new opportunities in nonlinear optics. Here we study the effects of the source incoherence in second harmonic generation. We derive and experimentally validate an analytic expression for the far-field coherence function at the second harmonic frequency, which is a generalization of the well-known van Cittert-Zernike theorem. Unlike the linear optics case, here the results depend not only on the source spatial distribution, but also on the nonlinear structure of the crystal. We further show, by performing Young's double-slit measurement, that the coherence at the second harmonic increases as the light propagates inside the nonlinear crystal. Our results establish nonlinearity as a degree of freedom for optical coherence, opening a new research area at the interface of statistical optics and nonlinear optics.
\end{abstract}

\maketitle

%\tableofcontents

\begin{figure}[htbp]
    \centering
    \includegraphics[width=0.45\textwidth]{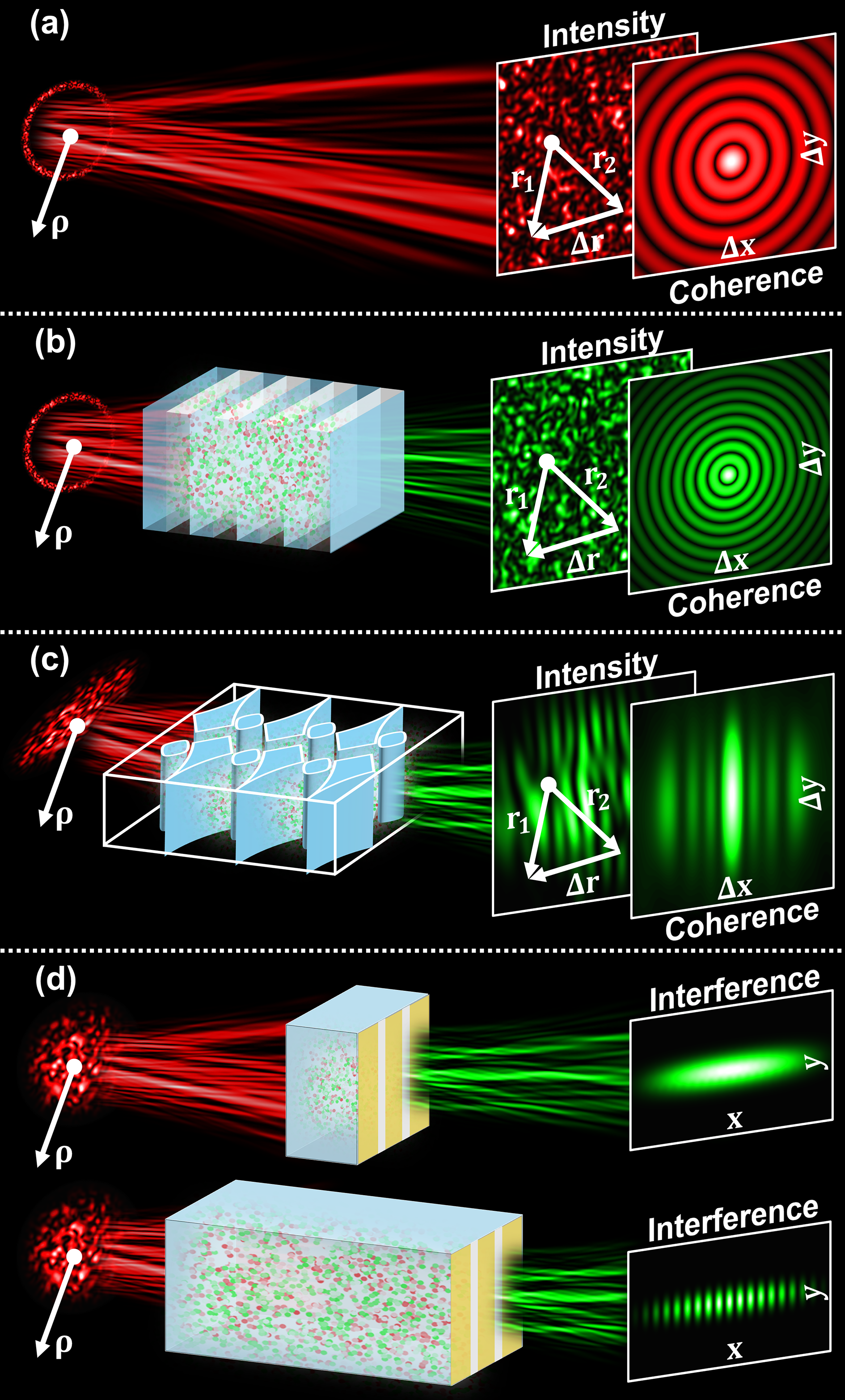}
    \caption{Formation of SOC in linear and nonlinear optics. Red and green denote the fundamental and SH beams, respectively. (a) Producing Bessel coherence at $\omega$ by incoherent annular sources in free space. (b) Producing Bessel coherence at $2\omega$ by incoherent annular sources with a conventional PPKTP crystal. (c) Producing Cartesian Laguerre-Gaussian coherence at $2\omega$ with an HG$_{03}$ nonlinear crystal. (d) Young's double-slit experiments proving the coherence growth at $2\omega$ as a function of the interaction length. Upper figure: 1 mm PPKTP crystal; Lower figure: 50 mm LBO crystal.}
    \label{f1}
\end{figure}

\emph{Introduction.}---Optical coherence, made experimentally visible through Young's interference experiment \cite{young1804bakerian}, has long quantified as a scalar measure of the ability of light to interfere, and was exploited in Michelson's stellar interferometer \cite{michelson1920application} to determine stellar angular sizes from fringe visibility. The van Cittert-Zernike (VCZ) theorem \cite{van1934wahrscheinliche,zernike1938concept} broadened this view by showing that the far-field first-order coherence function is given by the Fourier transform of an incoherent source. It therefore established coherence not merely as a scalar measure, but as a structured spatial distribution governed by propagation and source geometry. As illustrated in Fig. \ref{f1}(a), a fully incoherent annular source can generate Bessel-type structured optical coherence (SOC) in the far field through linear propagation. In this sense, the VCZ theorem identifies coherence as a structured degree of freedom of light, alongside amplitude, phase, and polarization.

This structure-based view was later extended beyond first-order coherence by Hanbury Brown-Twiss intensity interferometry \cite{brown1954lxxiv,brown1956correlation,brown1958interferometry}, which established $g^{(2)}$ and higher-order correlations as fundamental observables in astronomy \cite{brown1954lxxiv} and quantum optics \cite{glauber1963quantum}. Recent studies \cite{redding2011spatial,redding2012speckle,knitter2016coherence,cao2019complex,eliezer2022controlling,roques2024measuring,mor2026separating,koivurova2021coherence,baek2023phase} have further reinforced the structured nature of optical coherence. In complex lasers \cite{redding2011spatial,redding2012speckle,knitter2016coherence,cao2019complex,eliezer2022controlling}, partial spatial coherence is controlled through the superposition of multiple spatial modes rather than treated as a single global quantity. In integrated photonics \cite{roques2024measuring,mor2026separating}, self-configuring interferometric networks treat partial coherence as a diagonalizable matrix of spatial states. Metamaterials have also enabled direct control of SOC \cite{koivurova2021coherence}, allowing laser light to be tuned between nearly incoherent and fully coherent states. Together, these developments show that SOC is becoming a central concept for controlling and engineering light.

In nonlinear optical systems, the effect of SOC has been widely studied in a variety of contexts, but most often through its impact on other observables rather than through a direct study of the SOC itself. For example, in $\chi^{(3)}$ nonlinear media \cite{christodoulides1997incoherent,christodoulides1997theory,chen1998self,kip2000modulation,coskun2000bright,rotschild2008incoherent}, the Gaussian coherence in the Gaussian-Schell beams was shown to strongly influence the formation, propagation, and stability of incoherent spatial solitons. In spontaneous parametric down-conversion \cite{defienne2019spatially,Zhang:19}, it has been directly used as a resource for tailoring the spatial correlations of down-converted photons and their degree of entanglement. Work \cite{joobeur1996coherence,saleh2000duality} inspired by the VCZ theorem established a duality between partial coherence and partial entanglement, including a biphoton counterpart of the classical coherence. More recently, it was shown that pump beams with the twisted form of coherence can lead to a counterintuitive regime where the entanglement of the down-converted state increases as the pump becomes less coherent \cite{hutter2020boosting}. Yet only in rare cases \cite{agrawal1981second,waller2012phase,pang2025coherence} has the SOC itself been treated as the central object of study in a nonlinear medium. An early step \cite{agrawal1981second} in this direction was Agrawal's analytical treatment of $\chi^{(2)}$ frequency conversion, which describes the general form of coherence for the second harmonic (SH) explicitly in terms of the higher-order coherence of the fundamental. However, because coherence functions are inherently high-dimensional ($\geq 4$) and are further coupled through nonlinear wave mixing, the direct characterization and control of specific structured features of optical coherence in nonlinear media remain largely underdeveloped.

In this Letter, we study the effects of incoherent light in nonlinear optics. Specifically, we measure and analyze the effects of an incoherent pump on the coherence function of the SH signal. We derive an analytic expression that relates the source spatial function and the SH far-field coherence function, which is a generalization of the linear VCZ theorem. We show that the result depends not only on the source parameters, but also on the shape of the crystal nonlinearity. For the full nonlinear interactions in long crystals, characterizing the structured properties of the coherence in the SH presents a significant challenge due to the fact that the stochastic process in this regime is inherently nonstationary. Here, using Young’s double-slit experiment, we directly observe the accumulation of coherence along a long nonlinear crystal.

\emph{Nonlinear van Cittert-Zernike theorem.}---We began by considering the scenario illustrated in Fig. \ref{f1}(b), where a fully incoherent fundamental source with a shaped intensity profile is launched into a nonlinear crystal to achieve second harmonic generation. The far-field spatial coherence of the SH is described by the four-dimensional (4D) mutual coherence function \cite{Mandel_Wolf_1995}
\begin{equation}\label{farfieldcoherenceofSHG}
\begin{aligned}
\Gamma_{2\omega}^{\infty}(\mathbf{r_1}, \mathbf{r_2}) & = \left\langle [\mathrm{E}_{2\omega}^{\infty}(\mathbf{r_1})]^* [\mathrm{E}_{2\omega}^{\infty}(\mathbf{r_2})] \right\rangle \\
& = \left\langle [(\hat{\mathrm{F}}\hat{\mathrm{N}}\mathrm{E}_{\omega})(\boldsymbol{\mathbf{r}_1}) ]^* [(\hat{\mathrm{F}}\hat{\mathrm{N}}\mathrm{E}_{\omega})(\boldsymbol{\mathbf{r}_2}) ] \right\rangle 
\end{aligned}
\end{equation}
where ${\langle \cdot \rangle}$ represents the statistical average; $*$ represents the complex conjugate; $\mathrm{E}_{\omega}$ and $\mathrm{E}_{2\omega}^{\infty}$ are respectively the input fundamental beam at frequency $\omega$ and the far-field SH at $2\omega$; $\mathbf{r}_i = x_i\mathbf{\hat{x}} + y_i\mathbf{\hat{y}}$ $(i=1,2)$ is the transverse two-dimensional (2D) position vector in the far field; $\hat{\mathrm{F}}$ denotes the Fourier transform operator used to describe far-field propagation, defined as \cite{goodman2005introduction}: $(\hat{\mathrm{F}} \mathrm{U})(\mathbf{r}) = \int \mathrm{U}(\bm{\uprho}) \exp \left(- i k \mathbf{r} \cdot \bm{\uprho} / f \right) \mathrm{d}^2 \bm{\uprho}$ with the initial transverse 2D position vector $\bm{\uprho} = \xi\mathbf{\hat{x}} + \eta\mathbf{\hat{y}}$, the wave numbers $k$, and the focal length of the Fourier lens $f$; 

The operator $\hat{\mathrm{N}}$ maps the input fundamental to the coherent SH at the crystal exit plane. In general, $\hat{\mathrm{N}}\mathrm{E}_{\omega}$ is obtained by solving the nonlinear coupled-wave equations under the time-harmonic, slowly varying envelope approximation \cite{boyd2008nonlinear}, and no closed-form solution is available for fully nonlinear propagation. Under the weak-interaction approximation in a short crystal, however, the SH field at the exit plane reduces to \cite{arie2010periodic}: $(\hat{\mathrm{N}}\mathrm{E}_{\omega})\left(\bm{\uprho}\right) = \kappa \chi_{\perp}^{(2)}\left(\bm{\uprho}\right) \mathrm{E}_{\omega}^2\left(\bm{\uprho}\right)$,
where $\kappa$ is the nonlinear coupling coefficient and $\chi_{\perp}^{(2)}(\bm{\uprho})$ is the engineered transverse nonlinear susceptibility, including both amplitude and phase. Such a transverse nonlinear profile can be realized in ferroelectric crystals by electric field poling \cite{armstrong1962interactions,berger1998nonlinear}.

\begin{figure}[htbp]
    \centering
    \includegraphics[width=0.48\textwidth]{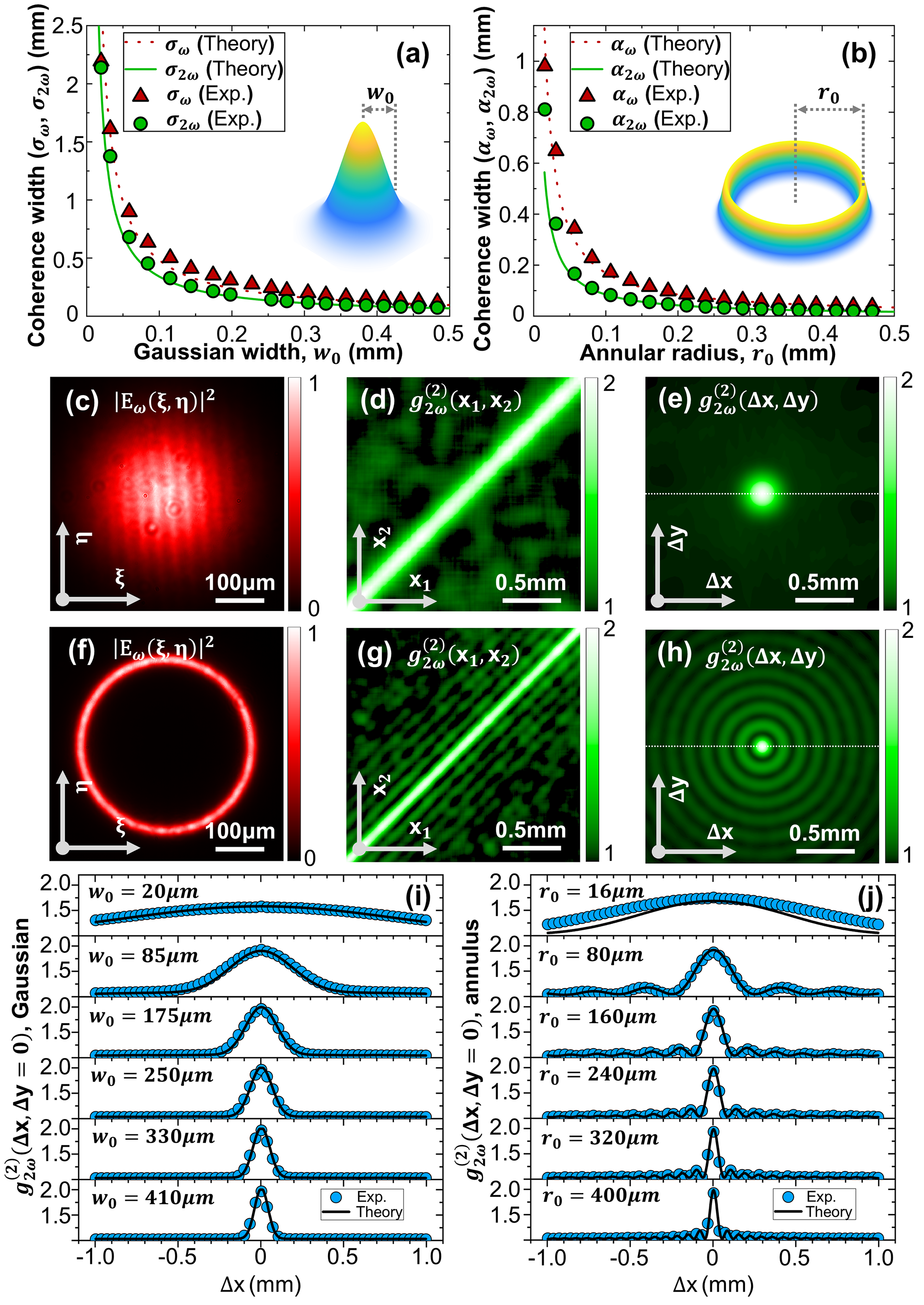}
    \caption{Gaussian and Bessel coherence in the fundamental and SH. (a) Gaussian coherence widths $(\sigma_\omega, \sigma_{2\omega})$ versus the Gaussian intensity width $w_0$. (b) Bessel coherence widths $(\alpha_\omega, \alpha_{2\omega})$ versus the ring radius $r_0$. (c) Gaussian intensity profile at $\omega$. (d),(e) Corresponding 1D and 2D Gaussian coherence at $2\omega$. (f) Annular intensity profile at $\omega$. (g),(h) Corresponding 1D and 2D Bessel coherence at $2\omega$. Theory-experiment comparisons are shown for Gaussian (i) and Bessel coherence (j) for various $w_0$ and $r_0$, respectively. Other parameters are $w_r = 20 $ $\upmu$m, $\sigma_\mu = 20$ $\upmu$m, and $f = 100$ mm.}
    \label{f2}
\end{figure}

Substituting this short-crystal solution and the Fourier propagation operator into into Eq. (\ref{farfieldcoherenceofSHG}), we derive the far-field coherence of the SH for an incoherent fundamental beam as the form,
\begin{equation}\label{nonlinearVCZ}
\begin{aligned}
\Gamma_{2\omega}^{\infty}(\Delta \mathbf{r}) \propto \int \left| \chi_{\perp}^{(2)}\left(\bm{\uprho}\right) \mathrm{E}_{\omega}^2\left(\bm{\uprho}\right) \right|^2 e^{-i k_{2} \Delta \mathbf{r} \cdot \bm{\uprho} / f} \, \mathrm{d}^2 \bm{\uprho},
\end{aligned}
\end{equation}
where $\Delta \mathbf{r} = \mathbf{r}_1 - \mathbf{r}_2$ is the transverse displacement vector between two transverse positions.

Equation (\ref{nonlinearVCZ}) implies that $\Gamma_{2\omega}^{\infty}$ is proportional to the Fourier transform of $\left| \chi_{\perp}^{(2)} (\bm{\uprho}) \mathrm{E}_{\omega}^2\right|^2$. It bears a resemblance in form to the VCZ theorem, which writes $
\Gamma_{\omega}^{\infty}(\Delta \mathbf{r}) \propto \int \left| \mathrm{E}_{\omega}\left(\bm{\uprho}\right) \right|^2 e^{-i k_{1} \Delta \mathbf{r} \cdot \bm{\uprho} / f} \, \mathrm{d}^2 \bm{\uprho}$. In this sense, Eq. (\ref{nonlinearVCZ}) constitutes a nonlinear VCZ theorem, stating that the spatially varying nonlinearity $\chi^{(2)}(\bm{\uprho})$ and the squared fundamental field $\mathrm{E}_{\omega}^2$ jointly modulate the far-field coherence of the SH. Importantly, this modulation depends only on the collective intensity of the product $\chi_{\perp}^{(2)} \mathrm{E}_{\omega}^2$, with all phase information removed from the modulation itself due to the stochastic process. Furthermore, as in linear optics, the resulting far-field coherence collapses from the 4D function defined over two arbitrary points $(\mathbf{r}_1, \mathbf{r}_2)$ in Eq. (\ref{farfieldcoherenceofSHG}), to a 2D function that depends only on their spatial separation $(\Delta \boldsymbol{\mathbf{r}})$ in Eq. (\ref{nonlinearVCZ}).

Throughout this Letter, we utilize the proposed nonlinear VCZ theorem to generate SOC at new frequencies. For readability, we have gathered the detailed derivations of both the theorem itself and the resulting SOC in Section A of End Matter.

\emph{Effect of pump distribution.}---To explore the nonlinear VCZ theorem, we first consider shaped incoherent fundamental beams propagating in a crystal with homogeneous transverse nonlinearity, $\chi_{\perp}^{(2)}(\bm{\uprho})=1$.  We analyze two types of spatially incoherent fundamental inputs: a Gaussian input, $\left|\mathrm{E}_{\omega}(\bm{\uprho})\right|^2 = \exp(-2\bm{\uprho}^2/w_0^2)$, and an annular input, $\left|\mathrm{E}_{\omega}(\bm{\uprho})\right|^2 = \exp[-2(|\bm{\uprho}|-r_0)^2/w_r^2]$. As predicted by Eq. (\ref{nonlinearVCZ}), these inputs generate, respectively, Gaussian far-field coherence, $\Gamma_{j}^{\infty}(\Delta \mathbf{r}) \propto \exp[-(\Delta \mathbf{r})^2/\sigma_j^2]$, and Bessel far-field coherence, $\Gamma_{j}^{\infty}(\Delta \mathbf{r}) \propto J_0(|\Delta \mathbf{r}|/\alpha_j)$, for both $j=\omega$ and $2\omega$. Moreover, owing to the intrinsically different propagation mechanisms of the two fields---linear propagation for the fundamental and nonlinear generation for the SH---their corresponding coherence widths are different, namely, $\sigma_\omega \left( w_0 \right) = \sqrt{2} \sigma_{2\omega}\left( w_0 \right)$ and $\alpha_\omega \left( r_0 \right) = 2 \alpha_{2\omega} \left( r_0 \right)$. Here, the Gaussian coherence width at the fundamental is $\sqrt{2}$ times that at the SH, whereas the Bessel coherence width is twice as large. This distinction arises because a Gaussian input is narrowed during second harmonic generation, from $w_0$ to $w_0/\sqrt{2}$, in addition to the wave-vector doubling from $k_1$ to $2k_1$. By contrast, the annular radius remains unchanged, so the Bessel width is reduced solely by the twofold increase in wave vector.

Experimentally, we use a phase-only spatial light modulator (SLM) illuminated at $1064.5~\mathrm{nm}$ to produce 1000 speckle realizations, $\left[\mathrm{E}_{\omega}(\bm{\uprho})\mathrm{T}^{(1)}(\bm{\uprho}), ..., \mathrm{E}_{\omega}(\bm{\uprho})\mathrm{T}^{(\rm{1000})}(\bm{\uprho}) \right]$, where $\mathrm{T}^{(\rm{n})}(\bm{\uprho})$ represents the ${\rm{n}}^{\rm{th}}$ frame of random complex transmission function with the statistical property $ \left\langle \mathrm{T}^{*}(\bm{\uprho_1})\mathrm{T}^{}(\bm{\uprho_2}) \right\rangle = e^{-( \bm{\uprho}_1 - \bm{\uprho}_2 )^2 / \sigma_\mu^2}$. These speckle realizations collectively form a spatially incoherent fundamental beam with the prescribed intensity profile, which are then launched into a 1-mm periodically poled $\mathrm{KTiOPO_4}$ (PPKTP) crystal to generate SH speckles at $532~\mathrm{nm}$. The far-field intensities of both fields are recorded frame by frame. Details of the experimental setup are provided in End Matter (Section B). From these intensity patterns, we evaluate the normalized intensity correlation function, $g^{(2)}\left(\Delta\mathbf{r}\right)
= \left\langle \int \mathrm{d}^2\mathbf{r}\,
I(\mathbf{r}) I(\mathbf{r}+\Delta\mathbf{r}) \right\rangle
 / \int \mathrm{d}^2\mathbf{r}\,
\langle I(\mathbf{r}) \rangle
\langle I(\mathbf{r}+\Delta\mathbf{r}) \rangle
= 1+\left|
\Gamma(\Delta\mathbf{r}) / \Gamma(0)
\right|^2$. Here, the second equality follows from the Siegert relation \cite{siegert1943fluctuations}, which allows the field correlation $\Gamma(\Delta\mathbf{r})$ to be inferred from the experimentally accessible intensity correlation $g^{(2)}(\Delta\mathbf{r})$. All subsequent characterizations of the SOC are therefore based on $g^{(2)}$.

Figure \ref{f2} presents simulations and measurements of the nonlinear generation of Gaussian and Bessel coherence. In the simulations, incoherent fundamental beams are represented by speckle ensembles, and each realization is propagated through second harmonic generation using the split-step Fourier method \cite{agrawal2013nonlinear}. Figures \ref{f2}(a) and \ref{f2}(b) show the Gaussian coherence widths $(\sigma_\omega,\sigma_{2\omega})$ versus the input width $w_0$, and the Bessel coherence widths $(\alpha_\omega,\alpha_{2\omega})$ versus the annular radius $r_0$, respectively. As shown here, the Gaussian and Bessel coherence widths scale as $1/w_0$ and $1/r_0$, respectively, as a consequence of far-field propagation. The measurements also confirm $\sigma_\omega=\sqrt{2}\sigma_{2\omega}$ and $\alpha_\omega=2\alpha_{2\omega}$.

\begin{figure*}[htbp]
    \centering
    \includegraphics[width=0.95\textwidth]{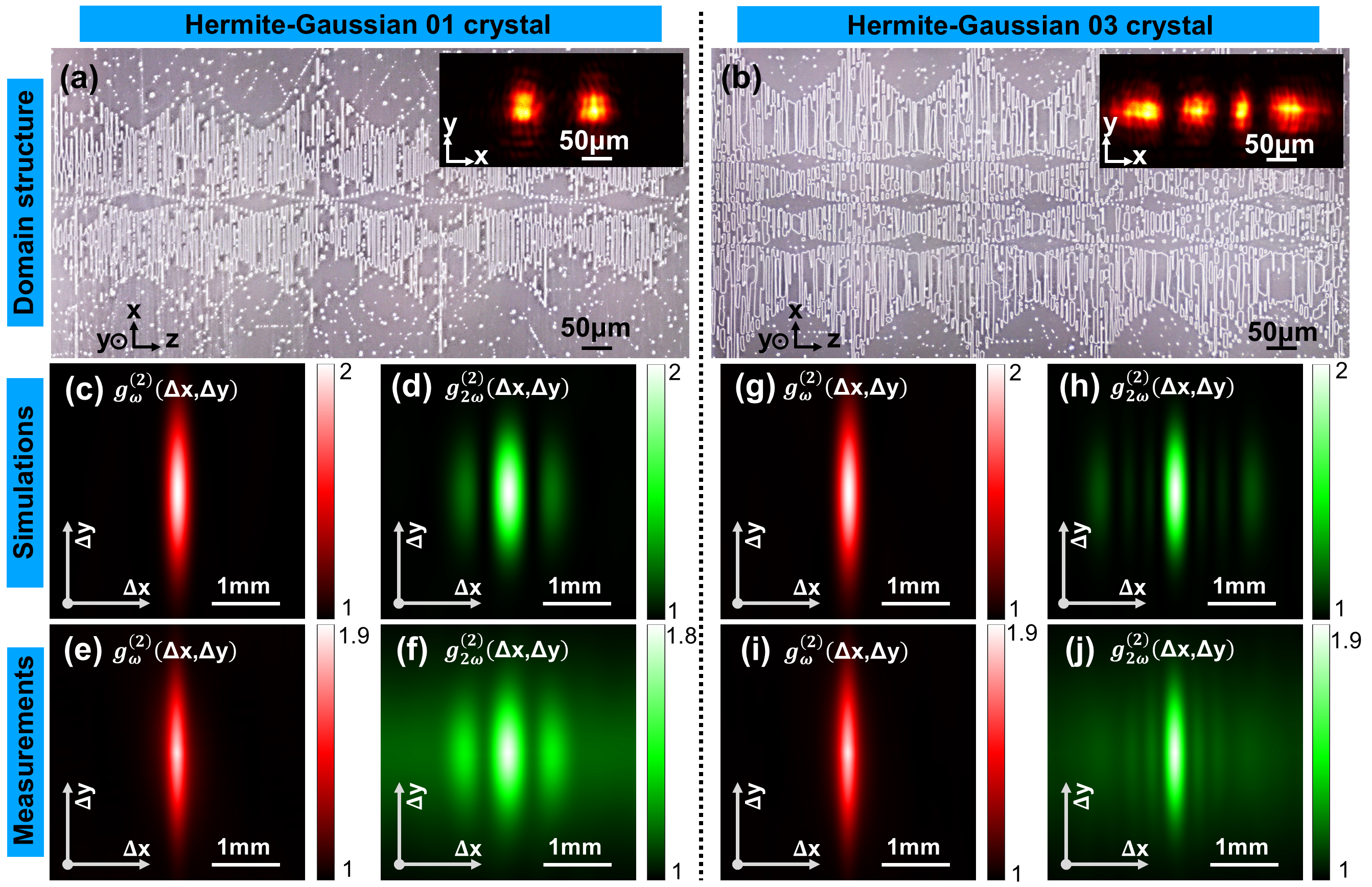}
    \caption{SOC generated using HG crystals. Microscope images of the HG$_{01}$ (a) and HG$_{03}$ (b) crystals. The crystals were selectively etched, in order to reveal the nonlinear modulation pattern \cite{trajtenberg2015on}. The insets show laser-based SH measurements of the crystals. (c)–(f) Simulations and measurements of the SOC of the fundamental beam (red) and the generated SH (green) after propagation through the HG$_{01}$ crystal. (g)–(j) Same as (c)–(f), but for the HG$_{03}$ crystal. The parameters are $w_{0x} = 475$ $\upmu$m, $w_{0y} = 70$ $\upmu$m, $w_{x} = 150$ $\upmu$m, $\sigma_{\mu} = 20$ $\upmu$m, and $f = 300$ mm.}
    \label{f3}
\end{figure*}

Figures \ref{f2}(c)-\ref{f2}(e) illustrate the process of pumping incoherent Gaussian fundamental beams [Fig. \ref{f2}(c)] to generate 1D [Fig. \ref{f2}(d)] and 2D [Fig. \ref{f2}(e)] Gaussian coherence in the SH, while Figs. \ref{f2}(f)-\ref{f2}(h) show the corresponding process for an incoherent annular input [Fig. \ref{f2}(f)], producing 1D [Fig. \ref{f2}(g)] and 2D [Fig. \ref{f2}(h)] Bessel coherence. For quantitative comparison, the cross sections $g^{(2)}(\Delta x, \Delta y=0)$ of the SOC extracted along the dashed lines in Figs. \ref{f2}(e) and \ref{f2}(h), are shown in Figs. \ref{f2}(i) and \ref{f2}(j) for various $w_0$ and $r_0$, respectively. The experimental results are in excellent agreement with theory, as evidenced in particular by the accurate reproduction of the Bessel side lobes in Fig. \ref{f2}(j).

\emph{Effect of engineered nonlinearity on shaping the coherence.}---We now extend the above discussion to the dependence of the engineered nonlinearity on the composition of the SOC. As shown in Fig. \ref{f1}(c), we consider an incoherent fundamental beam with elliptic Gaussian intensity profiles, one where $\left| \mathrm{E}_{\omega}\left(\bm{\uprho}\right)\right|^2 = e^{-2{\xi}^2 / w_{0x}^2 - 2{\eta}^2 / w_{0y}^2}$ incident on a periodically poled MgO doped stoichiometric lithium tantalate (PPSLT) crystal. The ferroelectric domains within this crystal are designed based on the computer generated holography methods \cite{Lee:79,trajtenberg2015on} and poled with a 1D Hermite-Gaussian (HG) spatial distribution, i.e., $\chi_{\perp}^{(2)}\left(\bm{\uprho}\right) = H_m\left( \sqrt{2} \xi / w_x\right) \exp \left(- {\xi}^2 / w_x^2 \right)$. As predicted by Eq. (\ref{nonlinearVCZ}), such a crystal yields the SH with the far-field SOC of the Cartesian Laguerre-Gaussian distribution (see End Matter), $\Gamma_{2\omega}^{\infty}(\Delta \mathbf{r}) \propto L_m \left[\left( {\Delta x}\right)^2 / W_x^2 \right]\exp \left[ - {\left( \Delta x \right)}^2 / 2 W_x^2 - {\left( \Delta y \right)}^2 / 2 W_y^2 \right]$, where $H_m$ and $L_m$ are respectively the $m$-th order Hermite and Laguerre polynomials \cite{gradshteyn2014table}; $W_x = 2 f / k_2 w_x$ and $W_y = 2 f / k_2 w_{0y}$ are the characteristic scales with $w_x $ the HG beam width.

Increasing the order of the HG crystal strengthens the oscillation in the SH coherence patterns, as reflected by the $m$ zeros of the resulting Laguerre polynomial. This $H_m \rightarrow L_m$ transformation is closely analogous to multi-beam interference. Because the random fluctuations of the incoherent fundamental wash out the phase information of the HG crystal, the SH coherence is governed solely by the crystal amplitude profile. Higher-order HG crystals therefore generate more spatially separated SH beams, for example two for HG$_{01}$ and four for HG$_{03}$. In the far field, these beams interfere to produce coherence patterns characteristic of multi-beam interference.

Figures \ref{f3}(a) and \ref{f3}(b) show microscope images of the HG$_{01}$ and HG$_{03}$ crystals, with insets confirming that a coherent pump generates HG beams in the SH. Figures \ref{f3}(c)-\ref{f3}(j) compare the far-field coherence of the fundamental and SH beams. For the HG$_{01}$ crystal [Figs. \ref{f3}(c)-\ref{f3}(f)], an incoherent elliptic fundamental beam produces elliptic fundamental coherence, with axes exchanged relative to the source intensity. In contrast, the SH coherence is determined by the Fourier transform of the HG$_{01}$ beam intensity in the crystal and is therefore independent of the fundamental intensity profile. This coherence decoupling marks a key distinction between the nonlinear VCZ theorem and its linear counterpart where the two fields share the same coherence pattern. Similar measurements for the HG$_{03}$ crystal [Figs. \ref{f3}(g)-\ref{f3}(j)] show unchanged fundamental coherence but SH coherence with additional interference fringes, confirming the analogy between HG-crystal-induced SOC and multi-beam interference.

\begin{figure}[htbp]
    \centering
    \includegraphics[width=0.48\textwidth]{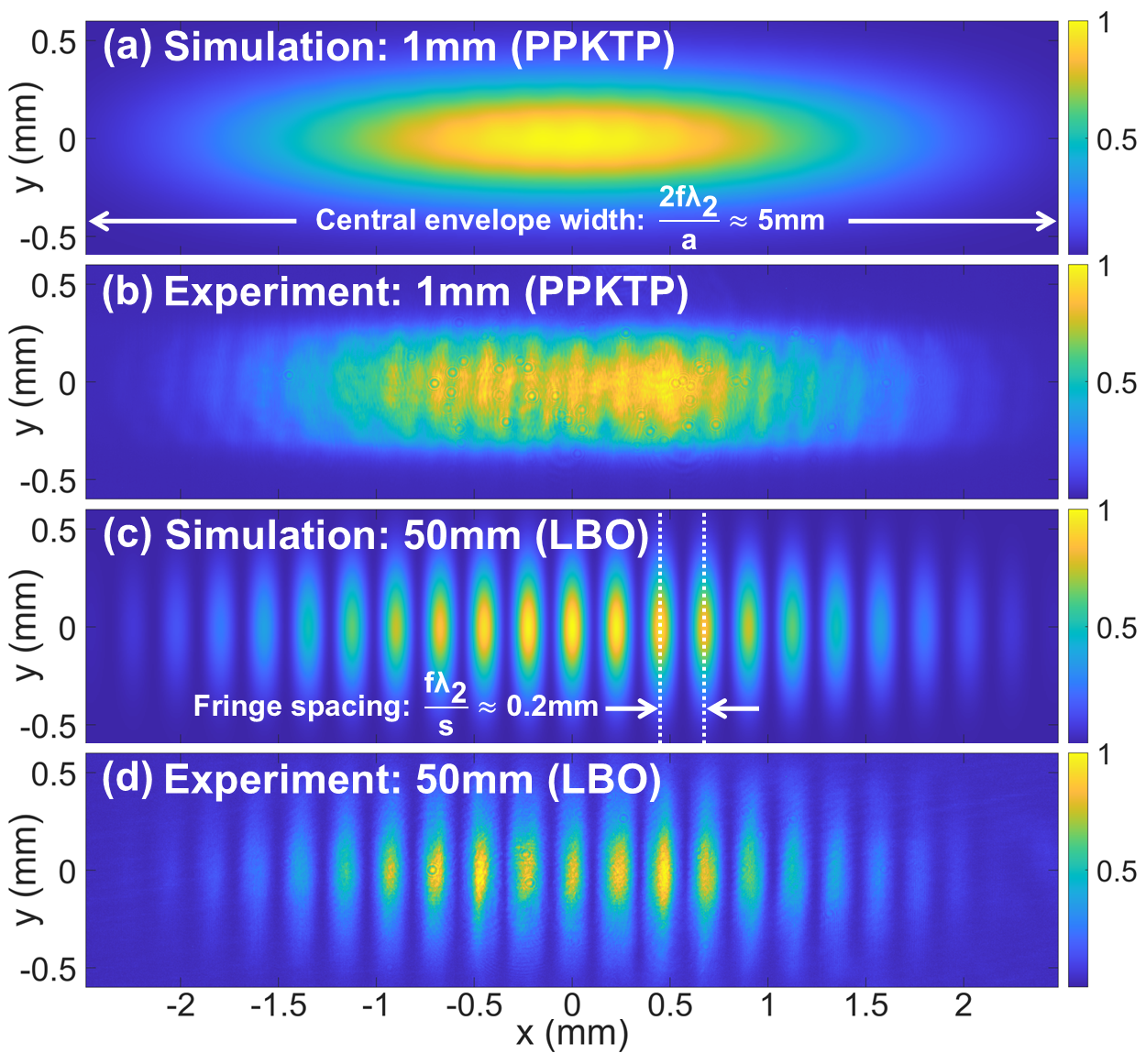}
    \caption{Young's double-slit experiment on the SH. Simulations (a) and experimental results (b) for the far-field double-slit interference fringes of the SH generated by a 1-mm PPKTP crystal. (c),(d) Same as (a),(b), but for a 50-mm LBO crystal. The beam parameters are $w_0 = 60$ $\upmu$m and $\sigma_\mu = 30$ $\upmu$m. The two slits have a width of $a = 5$ $\upmu$m each and separation of $s = 60$ $\upmu$m. $f$ = 25.4 mm.}
    \label{f4}
\end{figure}

\emph{Coherence evolution along the nonlinear crystal.}---The SOC construction above relies on the weak-interaction approximation for short nonlinear crystals. In long crystals, full nonlinear coupling generally leads to nonstationary SH coherence \cite{agrawal1981second}, so its complete description requires a 4D coherence function rather than the stationary 2D form used above. To probe this regime, we study the evolution of SH coherence during nonlinear propagation.

Using the same setup, we generated a Gaussian-Schell model fundamental beam at the crystal input having intensity width $w_0$ and coherence width $\sigma_\mu$ \cite{Mandel_Wolf_1995}, with $w_0=\sigma_\mu/2$ corresponding to an approximately incoherent input. As illustrated in Fig. \ref{f1}(d), we direct this fundamental beam into two different crystals: a short 1-mm PPKTP crystal and a long 50-mm $\mathrm{LiB_3O_5}$ (LBO) crystal. In addition, a double-slit mask is positioned at the output facet of each crystal using an optical 4$f$ system (see End Matter). The far-field double-slit pattern of the SH consists of interference fringes within the central diffraction envelope. The slit width $a$ sets the width of the central envelope, while the slit separation $s$ sets the fringe spacing. Since the two slits probe the field at points separated by $s$, the fringe visibility directly reflects the coherence strength between them.

The double-slit interference for the SH emerging from the crystals are shown in Figs. \ref{f4}(a)-\ref{f4}(d). As can be seen, the fringe visibility produced by the incoherent SH after passing through the short PPKTP crystal is nearly zero. For the long LBO crystal, however, using the identical input beam and double-slit mask at the output, the far-field interference fringes become clearly visible. The experimental measurements are in good agreement with the theoretical simulations. In linear optics, it is well known that coherence can be enhanced simply through free space propagation. Here, our results show that, in a nonlinear optical setting, an incoherent SH can similarly accumulate coherence during nonlinear propagation along the crystal.

\emph{Conclusions.}---We developed and experimentally validated a nonlinear optical extension of the VCZ theorem. Using this framework, we generated Gaussian and Bessel coherence in SH fields and showed that, in HG-patterned nonlinear crystals, the far-field SH coherence is governed by the spatial distribution of the nonlinearity. Young's double-slit measurements further revealed the accumulation of SH coherence during nonlinear propagation. These results open new opportunities for controlling stochastic optical process through nonlinear frequency conversion, including thermal emission and infrared incoherent imaging. The established correspondence between nonlinear structure and harmonic coherence may also provide a route for nondestructive characterization of internal domain patterns in nonlinear photonic materials.

More broadly, by linking a designed material response to a structured output correlation, our work may be extended to other nonlinear processes, such as $\chi^{(2)}$ sum-frequency generation and $\chi^{(3)}$ third-harmonic generation. Similar principles may also inspire coherence control in other wave systems, ranging from holographic gratings that couple incoherent light into surface plasmon polaritons \cite{dolev2012surface} to amplitude masks that shape the coherence of free-electron beams \cite{voloch2013generation,shiloh2015unveiling}.

\emph{Acknowledgments.}---We gratefully acknowledge the financial support provided by the Israel Science Foundation (Grants No. 969/22 and No. 3117/23). We thank the Tel Aviv University Center of Light-Matter Interaction for their support, and we are grateful to Dr. Inna Shekhtman for her assistance with the fabrication of the double-slit masks.

\bibliographystyle{apsrev4-2}
\bibliography{references}

@CONTROL{apsrev42Control,
  author="08",
  editor="1",
  pages="0",
  title="0",
  year="1"
}

@article{young1804bakerian,
  title={I. The Bakerian Lecture. Experiments and calculations relative to physical optics},
  author={Young, Thomas},
  journal={Philosophical transactions of the Royal Society of London},
  volume={94},
  pages={1--16},
  year={1804},
  publisher={The Royal Society London}
}

@article{michelson1920application,
  title={On the application of interference methods to astronomical measurements},
  author={Michelson, Albert Abraham},
  journal={Astrophys. J.},
  volume={51},
  pages={257},
  year={1920}
}

@article{van1934wahrscheinliche,
  title={Die wahrscheinliche Schwingungsverteilung in einer von einer Lichtquelle direkt oder mittels einer Linse beleuchteten Ebene},
  author={van Cittert, Pieter Hendrik},
  journal={Physica},
  volume={1},
  number={1-6},
  pages={201--210},
  year={1934},
  publisher={Elsevier}
}

@article{zernike1938concept,
  title={The concept of degree of coherence and its application to optical problems},
  author={Zernike, Frederik},
  journal={Physica},
  volume={5},
  number={8},
  pages={785--795},
  year={1938},
  publisher={Elsevier}
}

@article{brown1954lxxiv,
  title={LXXIV. A new type of interferometer for use in radio astronomy},
  author={Brown, R Hanbury and Twiss, Richard Q},
  journal={London, Edinburgh, Dublin Phil. Mag. J. Sci.},
  volume={45},
  number={366},
  pages={663--682},
  year={1954},
  publisher={Taylor \& Francis}
}

@article{brown1956correlation,
  title={Correlation between photons in two coherent beams of light},
  author={Brown, R Hanbury and Twiss, Richard Q},
  journal={Nature},
  volume={177},
  number={4497},
  pages={27--29},
  year={1956},
  publisher={Nature Publishing Group UK London}
}

@article{brown1958interferometry,
  title={Interferometry of the intensity fluctuations in light. II. An experimental test of the theory for partially coherent light},
  author={Brown, R Hanbury and Twiss, RQ},
  journal={Proc. R. Soc. A},
  volume={243},
  number={1234},
  pages={291--319},
  year={1958},
  publisher={The Royal Society London}
}

@article{glauber1963quantum,
 title={The quantum theory of optical coherence},
 author={Glauber, Roy J},
 journal={Phys. Rev.},
 volume={130},
 number={6},
 pages={2529},
 year={1963},
 publisher={APS}
}

@article{redding2011spatial,
  title={Spatial coherence of random laser emission},
  author={Redding, Brandon and Choma, Michael A and Cao, Hui},
  journal={Opt. Lett.},
  volume={36},
  number={17},
  pages={3404--3406},
  year={2011},
  publisher={Optical Society of America}
}

@article{redding2012speckle,
  title={Speckle-free laser imaging using random laser illumination},
  author={Redding, Brandon and Choma, Michael A and Cao, Hui},
  journal={Nat. photonics},
  volume={6},
  number={6},
  pages={355--359},
  year={2012},
  publisher={Nature Publishing Group}
}

@article{knitter2016coherence,
  title={Coherence switching of a degenerate VECSEL for multimodality imaging},
  author={Knitter, Sebastian and Liu, Changgeng and Redding, Brandon and Khokha, Mustafa K and Choma, Michael A and Cao, Hui},
  journal={Optica},
  volume={3},
  number={4},
  pages={403--406},
  year={2016},
  publisher={Optical Society of America}
}

@article{cao2019complex,
  title={Complex lasers with controllable coherence},
  author={Cao, Hui and Chriki, Ronen and Bittner, Stefan and Friesem, Asher A and Davidson, Nir},
  journal={Nat. Rev. Phys.},
  volume={1},
  number={2},
  pages={156--168},
  year={2019},
  publisher={Nature Publishing Group UK London}
}

@article{eliezer2022controlling,
  title = {Controlling Nonlinear Interaction in a Many-Mode Laser by Tuning Disorder},
  author = {Eliezer, Yaniv and Mahler, Simon and Friesem, Asher A. and Cao, Hui and Davidson, Nir},
  journal = {Phys. Rev. Lett.},
  volume = {128},
  issue = {14},
  pages = {143901},
  numpages = {5},
  year = {2022},
  month = {Apr},
  publisher = {American Physical Society},
  doi = {10.1103/PhysRevLett.128.143901},
  url = {https://link.aps.org/doi/10.1103/PhysRevLett.128.143901}
}

@article{roques2024measuring,
  title={Measuring, processing, and generating partially coherent light with self-configuring optics},
  author={Roques-Carmes, Charles and Fan, Shanhui and Miller, David AB},
  journal={Light: Sci. Appl.},
  volume={13},
  number={1},
  pages={260},
  year={2024},
  publisher={Nature Publishing Group UK London}
}

@article{mor2026separating,
  title={Separating partially coherent light},
  author={Mor, Paul-Alexis and Kroo, Anne R and Valdez, Carson G and {\v{S}}imi{\'c}, Marko and Karnieli, Aviv and Cavicchioli, Gabriele and Sun, Zhanghao and Grimaldi, Vittorio and Fan, Shanhui and Solgaard, Olav and others},
  journal={arXiv preprint arXiv:2603.15517},
  year={2026}
}

@article{koivurova2021coherence,
  title = {Coherence Switching with Metamaterials},
  author = {Koivurova, Matias and Hakala, Tommi K. and Turunen, Jari and Friberg, Ari T. and Caglayan, Humeyra and Ornigotti, Marco},
  journal = {Phys. Rev. Lett.},
  volume = {127},
  issue = {15},
  pages = {153902},
  numpages = {6},
  year = {2021},
  month = {Oct},
  publisher = {American Physical Society},
  doi = {10.1103/PhysRevLett.127.153902},
  url = {https://link.aps.org/doi/10.1103/PhysRevLett.127.153902}
}

@article{baek2023phase,
  title={Phase conjugation with spatially incoherent light in complex media},
  author={Baek, YoonSeok and De Aguiar, Hilton B and Gigan, Sylvain},
  journal={Nat. photonics},
  volume={17},
  number={12},
  pages={1114--1119},
  year={2023},
  publisher={Nature Publishing Group UK London}
}

@article{christodoulides1997incoherent,
  title={Incoherent spatial solitons in saturable nonlinear media},
  author={Christodoulides, DN and Coskun, TH and Joseph, RI},
  journal={Opt. Lett.},
  volume={22},
  number={14},
  pages={1080--1082},
  year={1997},
  publisher={Optical Society of America}
}

@article{christodoulides1997theory,
  title={Theory of incoherent self-focusing in biased photorefractive media},
  author={Christodoulides, Demetrios N and Coskun, Tamer H and Mitchell, Matthew and Segev, Mordechai},
  journal={Phys. Rev. Lett.},
  volume={78},
  number={4},
  pages={646},
  year={1997},
  publisher={APS}
}

@article{chen1998self,
  title={Self-trapping of dark incoherent light beams},
  author={Chen, Zhigang and Mitchell, Matthew and Segev, Mordechai and Coskun, Tamer H and Christodoulides, Demetrios N},
  journal={Science},
  volume={280},
  number={5365},
  pages={889--892},
  year={1998},
  publisher={American Association for the Advancement of Science}
}

@article{kip2000modulation,
  title={Modulation instability and pattern formation in spatially incoherent light beams},
  author={Kip, Detlef and Soljacic, Marin and Segev, Mordechai and Eugenieva, Eugenia and Christodoulides, Demetrios N},
  journal={Science},
  volume={290},
  number={5491},
  pages={495--498},
  year={2000},
  publisher={American Association for the Advancement of Science}
}

@article{coskun2000bright,
  title={Bright spatial solitons on a partially incoherent background},
  author={Coskun, Tamer H and Christodoulides, Demetrios N and Kim, Young-Rae and Chen, Zhigang and Soljacic, Marin and Segev, Mordechai},
  journal={Phys. Rev. Lett.},
  volume={84},
  number={11},
  pages={2374},
  year={2000},
  publisher={APS}
}

@article{rotschild2008incoherent,
  title={Incoherent spatial solitons in effectively instantaneous nonlinear media},
  author={Rotschild, Carmel and Schwartz, Tal and Cohen, Oren and Segev, Mordechai},
  journal={Nat. Photonics},
  volume={2},
  number={6},
  pages={371--376},
  year={2008},
  publisher={Nature Publishing Group UK London}
}

@article{defienne2019spatially,
  title = {Spatially entangled photon-pair generation using a partial spatially coherent pump beam},
  author = {Defienne, Hugo and Gigan, Sylvain},
  journal = {Phys. Rev. A},
  volume = {99},
  issue = {5},
  pages = {053831},
  numpages = {9},
  year = {2019},
  month = {May},
  publisher = {American Physical Society},
  doi = {10.1103/PhysRevA.99.053831},
  url = {https://link.aps.org/doi/10.1103/PhysRevA.99.053831}
}

@article{Zhang:19,
author = {Wuhong Zhang and Robert Fickler and Enno Giese and Lixiang Chen and Robert W. Boyd},
journal = {Opt. Express},
number = {15},
pages = {20745--20753},
publisher = {Optica Publishing Group},
title = {Influence of pump coherence on the generation of position-momentum entanglement in optical parametric down-conversion},
volume = {27},
month = {Jul},
year = {2019},
url = {https://opg.optica.org/oe/abstract.cfm?URI=oe-27-15-20745},
doi = {10.1364/OE.27.020745},
}

@article{joobeur1996coherence,
  title = {Coherence properties of entangled light beams generated by parametric down-conversion: Theory and experiment},
  author = {Joobeur, Adel and Saleh, Bahaa E. A. and Larchuk, Todd S. and Teich, Malvin C.},
  journal = {Phys. Rev. A},
  volume = {53},
  issue = {6},
  pages = {4360--4371},
  numpages = {0},
  year = {1996},
  month = {Jun},
  publisher = {American Physical Society},
  doi = {10.1103/PhysRevA.53.4360},
  url = {https://link.aps.org/doi/10.1103/PhysRevA.53.4360}
}

@article{saleh2000duality,
  title = {Duality between partial coherence and partial entanglement},
  author = {Saleh, Bahaa E. A. and Abouraddy, Ayman F. and Sergienko, Alexander V. and Teich, Malvin C.},
  journal = {Phys. Rev. A},
  volume = {62},
  issue = {4},
  pages = {043816},
  numpages = {15},
  year = {2000},
  month = {Sep},
  publisher = {American Physical Society},
  doi = {10.1103/PhysRevA.62.043816},
  url = {https://link.aps.org/doi/10.1103/PhysRevA.62.043816}
}

@article{hutter2020boosting,
  title = {Boosting Entanglement Generation in Down-Conversion with Incoherent Illumination},
  author = {Hutter, Lucas and Lima, G. and Walborn, S. P.},
  journal = {Phys. Rev. Lett.},
  volume = {125},
  issue = {19},
  pages = {193602},
  numpages = {5},
  year = {2020},
  month = {Nov},
  publisher = {American Physical Society},
  doi = {10.1103/PhysRevLett.125.193602},
  url = {https://link.aps.org/doi/10.1103/PhysRevLett.125.193602}
}

@article{agrawal1981second,
  title = {Second-harmonic generation with arbitrary pump-beam profiles},
  author = {Agrawal, G. P.},
  journal = {Phys. Rev. A},
  volume = {23},
  issue = {4},
  pages = {1863--1868},
  numpages = {0},
  year = {1981},
  month = {Apr},
  publisher = {American Physical Society},
  doi = {10.1103/PhysRevA.23.1863},
  url = {https://link.aps.org/doi/10.1103/PhysRevA.23.1863}
}

@article{waller2012phase,
  title={Phase-space measurement and coherence synthesis of optical beams},
  author={Waller, Laura and Situ, Guohai and Fleischer, Jason W},
  journal={Nat. Photonics},
  volume={6},
  number={7},
  pages={474--479},
  year={2012},
  publisher={Nature Publishing Group UK London}
}

@article{pang2025coherence,
  title={Coherence synthesis in nonlinear optics},
  author={Pang, Zihao and Arie, Ady},
  journal={Light: Sci. Appl.},
  volume={14},
  number={1},
  pages={101},
  year={2025},
  publisher={Nature Publishing Group UK London}
}

@book{Mandel_Wolf_1995,
author = {Mandel, Leonard and Wolf, Emil},
title = {Optical Coherence and Quantum Optics},
year = {1995},
publisher = {Cambridge University Press}
}

@book{goodman2005introduction,
author = {Goodman, Joseph W},
title = {Introduction to Fourier Optics},
year = {2005},
publisher = {Robert \& Company Publishers, Englewood, CO}
}

@book{boyd2008nonlinear,
author = {Boyd, Robert W.},
title = {Nonlinear Optics},
year = {2008},
isbn = {0123694701},
publisher = {Academic Press}
}

@article{arie2010periodic,
author = {Arie, A. and Voloch, N.},
title = {Periodic, quasi-periodic, and random quadratic nonlinear photonic crystals},
journal = {Laser Photonics Rev.},
volume = {4},
number = {3},
pages = {355-373},
doi = {https://doi.org/10.1002/lpor.200910006},
year = {2010}
}

@article{armstrong1962interactions,
  title = {Interactions between Light Waves in a Nonlinear Dielectric},
  author = {Armstrong, J. A. and Bloembergen, N. and Ducuing, J. and Pershan, P. S.},
  journal = {Phys. Rev.},
  volume = {127},
  issue = {6},
  pages = {1918--1939},
  numpages = {0},
  year = {1962},
  month = {Sep},
  publisher = {American Physical Society},
  doi = {10.1103/PhysRev.127.1918},
  url = {https://link.aps.org/doi/10.1103/PhysRev.127.1918}
}

@article{berger1998nonlinear,
  title = {Nonlinear Photonic Crystals},
  author = {Berger, V.},
  journal = {Phys. Rev. Lett.},
  volume = {81},
  issue = {19},
  pages = {4136--4139},
  numpages = {0},
  year = {1998},
  month = {Nov},
  publisher = {American Physical Society},
  doi = {10.1103/PhysRevLett.81.4136},
  url = {https://link.aps.org/doi/10.1103/PhysRevLett.81.4136}
}

@book{siegert1943fluctuations,
  title={On the fluctuations in signals returned by many independently moving scatterers},
  author={Siegert, AJF},
  year={1943},
  publisher={Radiation Laboratory, Massachusetts Institute of Technology, Cambridge, MA}
}

@book{agrawal2013nonlinear,
author = {Agrawal, Govind P},
title = {Nonlinear Fiber Optics},
year = {2013},
publisher = {Academic Press, Oxford},
}

@article{Lee:79,
author = {Wai-Hon Lee},
journal = {Appl. Opt.},
number = {21},
pages = {3661--3669},
publisher = {Optica Publishing Group},
title = {Binary computer-generated holograms},
volume = {18},
month = {Nov},
year = {1979},
url = {https://opg.optica.org/ao/abstract.cfm?URI=ao-18-21-3661},
doi = {10.1364/AO.18.003661},
}

@article{trajtenberg2015on,
author = {Trajtenberg-Mills, Sivan and Juwiler, Irit and Arie, Ady},
title = {On-axis shaping of second-harmonic beams},
journal = {Laser Photonics Rev.},
volume = {9},
number = {6},
pages = {L40-L44},
year = {2015}
}

@book{gradshteyn2014table,
  title={Table of integrals, series, and products},
  author={Gradshteyn, Izrail Solomonovich and Ryzhik, Iosif Moiseevich},
  year={2014},
  publisher={Academic press, New York}
}

@article{dolev2012surface,
  title = {Surface-Plasmon Holographic Beam Shaping},
  author = {Dolev, Ido and Epstein, Itai and Arie, Ady},
  journal = {Phys. Rev. Lett.},
  volume = {109},
  issue = {20},
  pages = {203903},
  numpages = {5},
  year = {2012},
  month = {Nov},
  publisher = {American Physical Society},
  doi = {10.1103/PhysRevLett.109.203903},
  url = {https://link.aps.org/doi/10.1103/PhysRevLett.109.203903}
}

@article{voloch2013generation,
  title={Generation of electron Airy beams},
  author={Voloch-Bloch, Noa and Lereah, Yossi and Lilach, Yigal and Gover, Avraham and Arie, Ady},
  journal={Nature},
  volume={494},
  number={7437},
  pages={331--335},
  year={2013},
  publisher={Nature Publishing Group UK London}
}

@article{shiloh2015unveiling,
  title = {Unveiling the Orbital Angular Momentum and Acceleration of Electron Beams},
  author = {Shiloh, Roy and Tsur, Yuval and Remez, Roei and Lereah, Yossi and Malomed, Boris A. and Shvedov, Vladlen and Hnatovsky, Cyril and Krolikowski, Wieslaw and Arie, Ady},
  journal = {Phys. Rev. Lett.},
  volume = {114},
  issue = {9},
  pages = {096102},
  numpages = {5},
  year = {2015},
  month = {Mar},
  publisher = {American Physical Society},
  doi = {10.1103/PhysRevLett.114.096102},
  url = {https://link.aps.org/doi/10.1103/PhysRevLett.114.096102}
}

@book{lebedev1972special,
  title={Special functions and their applications},
  author={N. N. Lebedev},
  year={1972},
  publisher={Dover, New York}
}

@article{durnin1987diffraction,
  title = {Diffraction-free beams},
  author = {Durnin, J. and Miceli, J. J. and Eberly, J. H.},
  journal = {Phys. Rev. Lett.},
  volume = {58},
  issue = {15},
  pages = {1499--1501},
  numpages = {0},
  year = {1987},
  month = {Apr},
  publisher = {American Physical Society},
  doi = {10.1103/PhysRevLett.58.1499},
  url = {https://link.aps.org/doi/10.1103/PhysRevLett.58.1499}
}

@article{bolduc2013exact,
author = {Eliot Bolduc and Nicolas Bent and Enrico Santamato and Ebrahim Karimi and Robert W. Boyd},
journal = {Opt. Lett.},
number = {18},
pages = {3546--3549},
publisher = {Optica Publishing Group},
title = {Exact solution to simultaneous intensity and phase encryption with a single phase-only hologram},
volume = {38},
month = {Sep},
year = {2013},
url = {https://opg.optica.org/ol/abstract.cfm?URI=ol-38-18-3546},
doi = {10.1364/OL.38.003546},
}

\section*{End Matter}

\appendix
\setcounter{section}{1}
\setcounter{equation}{0}
\renewcommand{\theequation}{A\arabic{equation}}

\emph{Section A: Deriving the nonlinear van Cittert-Zernike theorem and its induced coherence.}---Under the undepleted pump approximation, the paraxial propagation for the time-harmonic fundamental $\mathrm{E}_{\omega}$ and the SH beams $\mathrm{E}_{2\omega}$ is \cite{boyd2008nonlinear},
\begin{equation} \label{CWE}
\begin{aligned}
\nabla_{\perp}^2 \mathrm{E}_{2\omega} + 2ik_2 \frac{\partial \mathrm{E}_{2\omega} }{\partial z} = - \frac{\omega_2^2 \chi^{(2)}\left(\bm{\uprho}, z\right) }{c^2} \mathrm{E}_\omega^2 e^{i\Delta kz},
\end{aligned}
\end{equation}
where $\nabla_{\perp}^2 = \partial^2/\partial x^2 + \partial^2/\partial y^2$ is the transverse Laplacian operator, $\Delta k = 2k_1 - k_2$ is the phase mismatch, and $c$ is the speed of light in vacuum. The terms $k_{1,2}$ and $\omega_{1,2}$ ($\omega_2 = 2 \omega_1$) represent the wave numbers and angular frequencies of the fundamental and SH beams, respectively. The nonlinear coupling susceptibility tensor, $\chi^{(2)}$, assumes a constant value for birefringent phase matching. For quasi-phase matching, however, it acts as a spatial function describing the ferroelectric domain structure. Consider a nonlinear crystal with the following engineered nonlinearity:
\begin{equation}\label{chi2}
\begin{aligned}
\chi^{(2)}\left(\bm{\uprho}, z\right) & = d_{ij}  \mathrm{sign} \left\{ \cos\left[\frac{2\pi z}{\Lambda} + \phi(\bm{\uprho})\right] - \cos(\pi q(\bm{\uprho})) \right\}\\
&= d_{ij} \sum_{m=-\infty}^{\infty} \left[ \frac{\sin \pi m q(\bm{\uprho})}{\pi m} \right] e^{\left\{-i m \left[\frac{2\pi z}{\Lambda} + \phi(\bm{\uprho})\right]\right\}},
\end{aligned}
\end{equation}
where $d_{ij}$ is the effective nonlinear tensor element and $\Lambda=2\pi/\Delta k$ is the poling period. This spatial modulation follows the design principle of Lee's binary computer-generated holograms \cite{Lee:79}. Retaining only the first order and neglecting higher-order contributions \cite{trajtenberg2015on}, the effective nonlinear coefficient becomes $\chi^{(2)}(\bm{\uprho},z)=d_{ij}\chi_{\perp}^{(2)}(\bm{\uprho})e^{-i\Delta k z}$. The longitudinal factor $e^{-i\Delta k z}$ cancels the phase-mismatch term in Eq. (\ref{CWE}), while the complex transverse envelope $\chi_{\perp}^{(2)}(\bm{\uprho})$ determines the spatial amplitude and phase of the nonlinear crystal, with $|\chi_{\perp}^{(2)}(\bm{\uprho})|=\sin[\pi q(\bm{\uprho})]/\pi$ and $\arg[\chi_{\perp}^{(2)}(\bm{\uprho})]=\phi(\bm{\uprho})$.

For a short crystal whose length is much smaller than the Rayleigh ranges of the interacting beams, the diffraction term (i.e. the transverse Laplacian term) in Eq. (\ref{CWE}) can be neglected. Equation (\ref{CWE}) then reduces to \cite{arie2010periodic}
$\mathrm{d}\mathrm{E}_{2\omega}/\mathrm{d}z
=i\chi^{(2)}(\bm{\uprho},z)\omega_2^2\mathrm{E}_{\omega}^2/(2k_2c^2)$.
Substituting the first order contribution of $\chi^{(2)}(\bm{\uprho},z)$ gives the SH field at the crystal output facet,
\begin{equation}\label{nearfieldSH}
\begin{aligned}
\mathrm{E}_{2\omega}\left(\bm{\uprho}\right) = (\hat{\mathrm{N}}\mathrm{E}_{\omega})\left(\bm{\uprho}\right) = \kappa \chi_{\perp}^{(2)}\left(\bm{\uprho}\right) \mathrm{E}_{\omega}^2\left(\bm{\uprho}\right),
\end{aligned}
\end{equation}
where $\kappa=iLd_{ij}\omega_2^2/(2k_2c^2)$ is the nonlinear coupling coefficient and $L$ is the crystal length. The far-field SH field, obtained at the rear focal plane of a Fourier lens, is $\mathrm{E}_{2\omega}^{\infty}\left(\mathbf{r}\right) = (\hat{\mathrm{F}}\mathrm{E}_{2\omega})\left(\mathbf{r}\right) = - i k_2 / 2 \pi f \int \mathrm{E}_{2\omega}\left(\bm{\uprho}\right) e^{- i k_2 \mathbf{r} \cdot \bm{\uprho} / f } \mathrm{d}^2\bm{\uprho}$. The statistical properties of the SH field are thus described by the 4D mutual coherence function \cite{Mandel_Wolf_1995}
\begin{equation}\label{farfieldcoherenceofSHGappendix}
\begin{aligned}
\Gamma_{2\omega}^{\infty}(\mathbf{r_1}, \mathbf{r_2}) 
& = \frac{k_2^2}{4 \pi^2 f^2} \iint\left\langle [\mathrm{E}_{2\omega}\left(\bm{\uprho_1}\right)]^* [\mathrm{E}_{2\omega}\left(\bm{\uprho_2}\right)] \right\rangle \\
& \quad \times \exp \left[ - \frac{i k_2 \left( \mathbf{r_2} \cdot \bm{\uprho_2} - \mathbf{r_1} \cdot \bm{\uprho_1}\right)}{f} \right] \mathrm{d}^2\bm{\uprho_1} \mathrm{d}^2\bm{\uprho_2},
\end{aligned}
\end{equation}
where $*$ denotes the complex conjugate.

A partially spatially coherent fundamental field can be written as
$\mathrm{E}_{\omega}(\bm{\uprho})=
\mathrm{T}(\bm{\uprho})|\mathrm{E}_{\omega}(\bm{\uprho})|
\exp[i\phi_{\omega}(\bm{\uprho})]$,
where $\phi_{\omega}(\bm{\uprho})$ is the phase of the field and $\mathrm{T}(\bm{\uprho})$ is a fluctuating complex transmission function. In the fully incoherent limit, the transmission function is delta-correlated,
$\langle \mathrm{T}^*(\bm{\uprho}_1)
\mathrm{T}(\bm{\uprho}_2)\rangle
=\delta(\bm{\uprho}_1-\bm{\uprho}_2)$.
Substituting Eq. (\ref{nearfieldSH}), together with $\chi_{\perp}^{(2)}(\bm{\uprho})$ and $\mathrm{E}_{\omega}(\bm{\uprho})$, into Eq. (\ref{farfieldcoherenceofSHGappendix}) yields the far-field SH mutual coherence function
\begin{equation}\label{nonlinearVCZappendix}
\begin{aligned}
\Gamma_{2\omega}^{\infty}(\Delta \mathbf{r}) = \mathcal{I} \int \left| \chi_{\perp}^{(2)}\left(\bm{\uprho}\right) \mathrm{E}_{\omega}^2\left(\bm{\uprho}\right) \right|^2 e^{-i k_{2} \Delta \mathbf{r} \cdot \bm{\uprho} / f} \, \mathrm{d}^2 \bm{\uprho},
\end{aligned}
\end{equation}
where $\mathcal{I}=k_2^2|\kappa|^2/(4\pi^2f^2)$. The delta correlation sets $\bm{\uprho}_1=\bm{\uprho}_2=\bm{\uprho}$ in the integral, and the multiplicative property of the modulus gives
$|\chi_{\perp}^{(2)}||\mathrm{E}_{\omega}^2|
=|\chi_{\perp}^{(2)}\mathrm{E}_{\omega}^2|$.

\emph{Case 1: Incoherent Gaussian pump and homogeneous nonlinearity.}
For $|\mathrm{E}_{\omega}(\bm{\uprho})|^2=\exp(-2\bm{\uprho}^2/w_0^2)$ and $\chi_{\perp}^{(2)}(\bm{\uprho})=1$, Eq. (\ref{nonlinearVCZappendix}) gives the Gaussian coherence of the SH beam. The corresponding Gaussian coherence for the fundamental follows from the linear VCZ theorem \cite{Mandel_Wolf_1995}. Explicitly,
\begin{equation}\label{gaussiancoherenceSHFB}
\begin{aligned}
\Gamma_{2\omega}^{\infty}(\Delta \mathbf{r}) = \mathcal{C}_{2\omega}e^{-\left( \Delta \mathbf{r} \right)^2 / \sigma_{2\omega}^2},\quad \Gamma_{\omega}^{\infty}(\Delta \mathbf{r}) = \mathcal{C}_{\omega}e^{-\left( \Delta \mathbf{r} \right)^2 / \sigma_{\omega}^2}
\end{aligned}
\end{equation}
with normalization constants $\mathcal{C}_{\omega} = k_2^2 w_0^2 |\kappa|^2 / 32 \pi f^2 = \mathcal{C}_{2\omega} / 2$ and Gaussian coherence widths $\sigma_{\omega} = 4 \sqrt{2} f / k_2 w_0 = \sqrt{2}\sigma_{2\omega}$.

\emph{Case 2: Incoherent annular pump and homogeneous nonlinearity.}
For $|\mathrm{E}_{\omega}(\bm{\uprho})|^2
=\exp[-2(|\bm{\uprho}|-r_0)^2/w_r^2]$
and $\chi_{\perp}^{(2)}(\bm{\uprho})=1$, we evaluate Eq. (\ref{nonlinearVCZappendix}) in polar coordinates:
$\xi=|\bm{\uprho}|\cos\varphi$,
$\eta=|\bm{\uprho}|\sin\varphi$,
$\Delta x=|\Delta\mathbf{r}|\cos\theta$, and
$\Delta y=|\Delta\mathbf{r}|\sin\theta$.
Thus,
$\mathrm{d}^2\bm{\uprho}=|\bm{\uprho}|\mathrm{d}|\bm{\uprho}|\mathrm{d}\varphi$
and
$\Delta\mathbf{r}\cdot\bm{\uprho}
=|\Delta\mathbf{r}||\bm{\uprho}|\cos(\varphi-\theta)$.
The SH coherence becomes
\begin{equation}
\begin{aligned}
\Gamma_{2\omega}^{\infty}(\Delta \mathbf{r})&=\mathcal{I}\int_0^\infty \int_0^{2\pi}|\bm{\uprho}| \exp\left[-\frac{4(|\bm{\uprho}|-r_0)^2}{w_r^2}\right] \\
& \quad \times \exp\left[-i\frac{k_2|\Delta \mathbf{r}||\bm{\uprho}|}{f}\cos(\varphi - \theta)\right]\mathrm{d}|\bm{\uprho}|\mathrm{d}\varphi .
\end{aligned}
\end{equation}
Using $\int_0^{2\pi}\exp(-ia\cos\phi)\mathrm{d}\phi=2\pi J_0(a)$ \cite{lebedev1972special}, we obtain the Bessel-type SH coherence function,
\begin{equation}\label{besselSHcoherence}
\begin{aligned}
\Gamma_{2\omega}^{\infty}(\Delta \mathbf{r})&=2\pi \mathcal{I}J_0\left(\frac{ |\Delta \mathbf{r}| }{ \alpha_{2\omega}}\right) \int_0^\infty|\bm{\uprho}| e^{-4(|\bm{\uprho}| - r_0)^2 / w_r^2} \mathrm{d}|\bm{\uprho}|,
\end{aligned}
\end{equation}
where $\alpha_{2\omega}=f/(k_2r_0)$ is the Bessel coherence width. In the thin-annulus limit, $w_r\ll r_0$, the radial weight is localized near $|\bm{\uprho}|=r_0$ \cite{durnin1987diffraction}. The replacement
$J_0(k_2|\bm{\uprho}||\Delta\mathbf{r}|/f)
\approx J_0(k_2r_0|\Delta\mathbf{r}|/f)$ is thus used in Eq. (\ref{besselSHcoherence}). The remaining radial integral contributes only a constant amplitude factor. The Bessel coherence of the fundamental beam follows analogously from the linear VCZ theorem. Since $k_2=2k_1$, the Bessel coherence widths satisfy $\alpha_{\omega}=2\alpha_{2\omega}$.

\emph{Case 3: Incoherent elliptic pump and HG nonlinearity.} For $\left| \mathrm{E}_{\omega}\left(\bm{\uprho}\right)\right|^2 = e^{-2{\xi}^2 / w_{0x}^2 - 2{\eta}^2 / w_{0y}^2}$ and $\chi_{\perp}^{(2)}\left(\bm{\uprho}\right) = H_m\left( \sqrt{2} \xi / w_x\right) \exp \left(- {\xi}^2 / w_x^2 \right)$, where $H_m$ is the $m$-th order Hermite polynomials \cite{gradshteyn2014table}, Eq. (\ref{nonlinearVCZappendix}) writes
\begin{equation}\label{HGinducedcoherence}
\begin{aligned}
\Gamma_{2\omega}^{\infty}(\Delta x) = \int_{-\infty}^{\infty} H_m^2\left(\frac{\sqrt{2}\xi}{w_x}\right) \exp\left(-\frac{2\xi^2}{w_x^2}-i \frac{k_2 \Delta x}{f} \xi\right) d\xi.
\end{aligned}
\end{equation}
Here, we assume the elliptic fundamental beam is much broader along $x$ than the HG beam in the crystal, i.e., $w_{0x} \gg w_x$. The factor $e^{-2\xi^2/w_{0x}^2}$ is thus treated as effectively constant and omitted from the integral. Along $y$, the SH coherence reduces to a Gaussian integral, yielding $e^{-(\Delta y)^2/2W_y^2}$ with $W_y = 2f/k_2 w_{0y}$.

To evaluate Eq. (\ref{HGinducedcoherence}) analytically, we apply the dimensionless spatial coordinate $v = \sqrt{2}\xi / w_x$, which yields $d\xi = w_x dv / \sqrt{2}$. Substituting this into the phase term gives $-i (k_2 w_x \Delta x / \sqrt{2} f) v$. By defining $K = k_2 w_x \Delta x / \sqrt{2} f $, Eq. (\ref{HGinducedcoherence}) scales to $\Gamma_{2\omega}^{\infty}(\Delta x) = w_x / \sqrt{2} \int_{-\infty}^{\infty} H_m^2(v) e^{-v^2} e^{-i K v} dv$. To solve the core integral $\mathcal{J}_m(K) = \int_{-\infty}^{\infty} H_m^2(v) e^{-v^2} e^{-i K v} dv$, we construct the bilinear generating function $G(t, u) = \sum_{m,n=0}^{\infty} \frac{t^m u^n}{m! n!} \int_{-\infty}^{\infty} H_m(v) H_n(v) e^{-v^2} e^{-i K v} dv$. $\mathcal{J}_m(K)$ is exactly the cross-term integral associated with the diagonal $m=n$ in $G(t, u)$. Utilizing the standard exponential generating function for Hermite polynomials \cite{lebedev1972special}, $\sum_{m=0}^{\infty} H_m(v) \frac{t^m}{m!} = e^{2vt - t^2}$, the integral simplifies via completing the square to the closed form $G(t, u) = \sqrt{\pi} e^{-K^2/4} e^{-iKt} e^{-iKu} e^{2tu}$. The diagonal contribution, $m=n$, is obtained by extracting the coefficient of $t^m u^m / (m!)^2$ in the Taylor expansion, yielding $\mathcal{J}_m(K) = \sqrt{\pi} e^{-K^2/4} 2^m m! \sum_{s=0}^m \binom{m}{s} \frac{1}{s!} \left(-\frac{K^2}{2}\right)^s$. Recognizing that this sum is exactly the series expansion of the Laguerre polynomial $L_m(K^2/2)$ \cite{gradshteyn2014table}, we reach $\mathcal{J}_m(K) = \sqrt{\pi} 2^m m! L_m(K^2/2) e^{-K^2/4}$.

Finally, substituting substitute $K = k_2 w_x \Delta x / \sqrt{2} f$ back into $\mathcal{J}_m(K)$, we obtain the analytical coherence of the SH generated by the HG crystal along the $x$-axis
\begin{equation}
\begin{aligned}
\Gamma_{2\omega}^{\infty}(\Delta x)  = \frac{w_x \sqrt{\pi}}{\sqrt{2}} 2^m m! L_m \left[ \frac{\left({\Delta x}\right)^2}{W_x^2}  \right] \exp \left[- \frac{\left({\Delta x}\right)^2 }{2W_x^2} \right],
\end{aligned}
\end{equation}
where $W_x = 2 f / k_2 w_x$ is the characteristic width.

\setcounter{section}{2}
\setcounter{equation}{0}
\renewcommand{\theequation}{B\arabic{equation}}

\begin{figure}[htbp]
    \centering
    \includegraphics[width=0.47\textwidth]{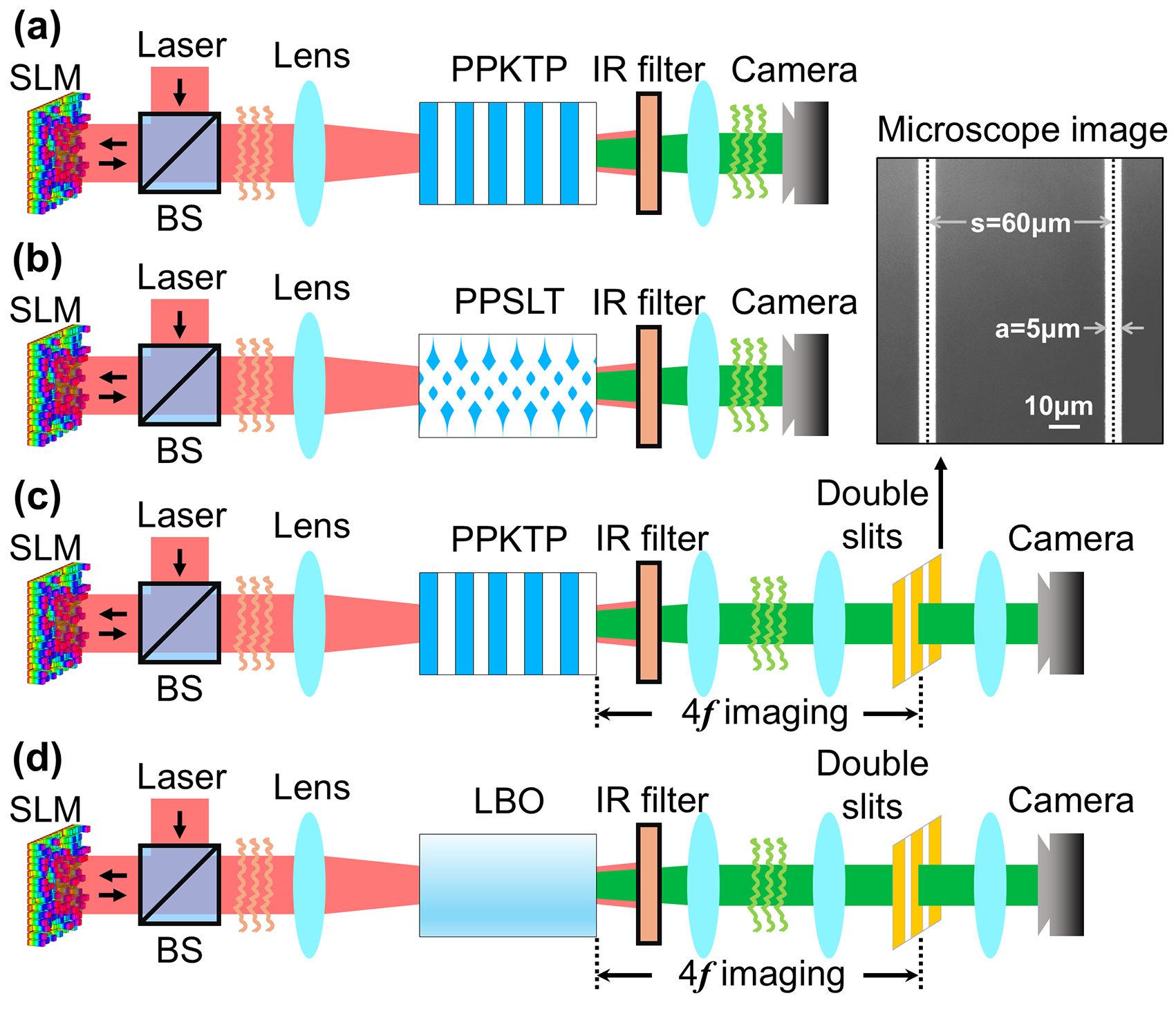}
    \caption{Schematics of the experimental setups.}
    \label{f5}
\end{figure}

\emph{Section B: Experimental setup.}---Throughout this Letter, we use three nonlinear crystals to perform the experiments shown in Fig.~\ref{f5}. In all experiments, a collimated, linearly polarized Nd:YAG Q-switched laser beam ($\uplambda = 1064.5$nm with pulse energy of 50 $\upmu$J, pulse width of 1 ns, and repetition rate of 1 kHz) illuminates the SLM (HoloEye PLUTO-2.1-NIR-133), which sequentially displays off-axis holograms to generate 1000 speckle-field realizations. These realizations form the desired incoherent fundamental beam at the crystal input facet. The holograms are generated by simulating the far-field scattering of a coherent laser and encoding the resulting scattered fields as phase-only holograms~\cite{bolduc2013exact}.

In the experiments studying the effect of the fundamental intensity profile, shown in Fig.~\ref{f5}(a), SH speckle fields are generated in a PPKTP crystal 
\((1\times2\times1~\mathrm{mm}^3, L\times W\times H)\)
with a \(9~\upmu\mathrm{m}\) poling period at \(32^\circ\mathrm{C}\). The crystal exit facet is placed at the front focal plane of a Fourier lens \((f=100~\mathrm{mm})\), and the speckle intensity is recorded at the back focal plane. In the experiments studying the effect of engineered nonlinearity, shown in Fig.~\ref{f5}(b), we use a PPSLT crystal 
\((1\times 1.4\times 0.5~\mathrm{mm}^3)\) with a \(7.9~\upmu\mathrm{m}\) poling period at \(95^\circ\mathrm{C}\) \cite{trajtenberg2015on}, containing HG-beam-shaped ferroelectric domains. The SH speckle intensity is measured at the back focal plane of a lens \((f=300~\mathrm{mm})\). In the experiments probing the evolution of SH coherence during nonlinear propagation, shown in Figs.~\ref{f5}(c) and \ref{f5}(d), we use the same PPKTP crystal and an LBO crystal 
\((50\times 4\times 4~\mathrm{mm}^3)\)
at \(148.5^\circ\mathrm{C}\), where the second harmonic process was based on type I non-critical birerfingent phase matching. The SH output from the crystal exit facet is imaged onto a double-slit mask using an optical \(4f\) system. The double-slit mask shown in the inset is fabricated by direct laser writing lithography (Heidelberg-Instrument DWL 66fs) of a chromium-coated quartz substrate, with each slit having a width $a = 5$ $\upmu$m and a separation $s = 60$ $\upmu$m between the two slits. A lens \((f=25.4~\mathrm{mm})\) then produces the far-field interference pattern of the SH beam transmitted through the mask.


\end{document}